\documentclass[aps,pra,twocolumn,10pt,showpacs,superscriptaddress,amsmath,amssymb]{revtex4-1}
\usepackage{amsmath,amssymb,bm}
\usepackage{graphicx}
\usepackage{epstopdf}
\usepackage{latexsym}
\usepackage{subfigure}
\usepackage[usenames,dvipsnames]{color}
\usepackage{hyperref}
\usepackage{natbib}
\usepackage{moreverb}
\begin{document}
\title{Demagnetization dynamics of non-interacting trapped fermions}
\date{\today}

\author{Andrew P. Koller}
\affiliation{JILA, NIST, Department of Physics, University of Colorado, 440 UCB, Boulder, CO 80309, USA}
\author{Joshua Mundinger}
\affiliation{Department of Mathematics and Statistics, Swarthmore College, 500 College Avenue, Swarthmore, PA 19081}
\author{Michael L. Wall}
\affiliation{JILA, NIST, Department of Physics, University of Colorado, 440 UCB, Boulder, CO 80309, USA}
\author{Ana Maria Rey}
\affiliation{JILA, NIST, Department of Physics, University of Colorado, 440 UCB, Boulder, CO 80309, USA}

\begin{abstract}
Motivated by several experimental efforts to understand spin diffusion and  transport in  ultracold fermionic gases, we study the spin dynamics of  initially  spin-polarized  ensembles of harmonically trapped non-interacting spin-1/2 fermionic atoms, subjected to a magnetic field gradient.  We obtain simple analytic expressions for spin observables in the presence of both constant and linear  magnetic field gradients, with and without a spin-echo pulse, and  at zero and finite temperatures. The analysis shows the relevance of spin-motional coupling in the non-interacting regime where  the demagnetization decay rate at short times  can  be  faster than the experimentally measured rates  in  the strongly interacting regime under similar trapping conditions. Our calculations also show that particle motion limits the ability of  a spin-echo pulse  to remove the effect of magnetic field inhomogeneity, and that a spin-echo pulse can instead lead to an increased decay of magnetization at times comparable to the trapping period.
\end{abstract}
\pacs{03.75.Ss, 05.30.Fk, 47.70.Nd, 67.85.Lm, 75.40.Gb}

\maketitle

\section{Introduction}
Understanding spin transport in quantum systems is  central to
many fields of physics and can have important applications for the development of quantum technologies. Recently there has been much experimental progress studying spin diffusion \cite{Koschorreck2013,Bardon14,Trotzky15,Natu2009, ebling11, Bruun2011, Piechon2009, xu2015, Goulko2013} and spin segregation dynamics  (time-dependent separation of the  spatial distributions of the spin components) \cite{Lewandowski2002, Du2008, Oktel2002, Fuchs2002, Williams2002, Bradley2002} in trapped atomic gases. A typical experimental protocol consists of preparing  a transversely spin-polarized gas and then observing the spin relaxation dynamics under the influence of a magnetic field gradient. For instance, Ref. \cite{Bardon14} measured the demagnetization timescale in a constant magnetic field  gradient and determined  the spin diffusion constant. Similarly, Ref. \cite{Du2008} measured the segregation of the spin populations in a magnetic field gradient with linear spatial dependence. Although the goal of the experiments is to understand the  many-body interacting spin dynamics, it is important to have a clear understanding of the non-interacting physics and how the spin-motion coupling alone affects the spin demagnetization. A thorough analysis  of the non-interacting  system will help provide the foundation  necessary for understanding the  complex spin dynamics induced by the interplay between interactions and motional effects.

 Here, we provide analytic expressions for the demagnetization exhibited by a harmonically trapped  and non-interacting spin-1/2 Fermi gas at both zero and finite temperatures. We study the spin dynamics  with and without a spin-echo pulse and in the presence of  magnetic field gradients with constant or linear spatial dependence.  The spin dynamics is oscillatory, and
 depends on details of the differential motion of the spin components.  In the case of a constant gradient (magnetic field that varies linearly with position),  the atoms oscillate at the trap frequency but in opposite directions along the gradient depending on their spin projection. We show that this periodic motion  gives rise to a fast demagnetization but not to a net spin rotation.  In the case of a linear gradient (magnetic field that varies quadratically with position), the spin components  breathe at different rates and with different magnitudes, the dynamics is not periodic with the trap frequency, and  the spin dynamics involves both a demagnetization and a net rotation of the collective spin.

  In the non-interacting limit our analysis reveals that the  transverse magnetization decays with a rate  that grows with increasing particle   number and temperature. The observed fast  demagnetization rate at short times reflects the fact that samples with a large number of fermions  occupy high harmonic oscillator modes that are widely spread across the trap and experience strong gradients. We also find that the spin-motion coupling  cannot be removed by a spin-echo pulse, and such a pulse can instead lead to an enhanced decay rate  of the magnetization of the gas. In unitary Fermi gas experiments this fast motion-induced demagnetization is suppressed at short times by interactions which instead lead to a diffusive decay of the magnetization. 

  This paper is organized as follows: In Sec.~\ref{sec1} we present and discuss the spectroscopic protocol under consideration. In Sec.~\ref{sec2} and Sec.~\ref{sec3} we present the constant and linear gradient cases, respectively. In each of these sections we derive analytic expressions for the single particle dynamics and then formulate the spin dynamics of the many-particle system at finite temperature in terms of these expressions.  We derive  expressions for the dynamics both in the presence and  absence of a spin-echo pulse and discuss experimental considerations. Finally, in Sec.\ref{sec4}, we finish with an outlook and conclusions.

\section{Spectroscopic Protocol}\label{sec1}
We begin by considering a single spin-1/2 particle of mass $m$ confined in a one-dimensional harmonic trap with trapping frequency $\omega$ -- results for ensembles of atoms are later calculated from sums over single particle dynamics. The spin dynamics is probed using Ramsey spectroscopy: at $t=0$ the particle is prepared in an eigenstate of the harmonic trap and with spin $\left|\downarrow\right\rangle$. The spin of the particle is then rotated about the $X$-axis by applying  a resonant pulse with
area $\theta_1$. Next, a position-dependent magnetic field $B_Z(x)$ pointing along $Z$ is suddenly turned on, and the particle then evolves freely in the presence of the magnetic field for a dark time $t$, after which spin observables are  measured.

Here, we focus on the case of  Ramsey spectroscopy with an initial pulse area  $\theta_1=\pi/2$ that rotates the initial state to point along $Y$ (transverse direction).  We  also assume that the pulse has zero detuning from the atomic transition ($\delta=0$).
However, the results that follow are easily generalized to arbitrary $\theta_1$ and finite detuning.

 During the dark time  the  particle  feels a potential
\begin{eqnarray}
\hat{V}(\bf x)&=& \hat{V}(x)+ \hat V(y,z) \nonumber \\
\hat{V}(x) &=&  \frac{1}{2}m\omega^2  x^2 
+ \Delta \mu B_Z(x) \hat \sigma^Z,
\end{eqnarray}
where $\Delta\mu$ is the differential magnetic moment between the two spin states and $\hat \sigma^Z$ is the usual Pauli operator.  Along the $y$ and $z$ directions the potential is assumed to be spin-independent. The dark time evolution of the state of the particle can be written as
\begin{eqnarray}
|\Psi ({\bf x},t)\rangle=\psi_{\uparrow}({\bf x},t)|\uparrow \rangle + \psi_{\downarrow}({\bf x},t)|\downarrow \rangle.
\end{eqnarray} Due to the separability and the spin independence of the $\hat V(y,z)$ potential, if the system is prepared in an eigenstate of $y$ and $z$, it will remain in that eigenstate and thus the dynamics is effectively one-dimensional. Because of this, we will restrict our analysis to the $x$ dimension.

We focus on the collective observable $\hat S^+ = \hat S^X + i\hat S^Y$, where $\hat S^\alpha$ are the spin angular momentum operators. The expectation value of $\hat S^+$  takes the form of an integrated density
\begin{eqnarray}
\langle \hat S^+\rangle  &=&\frac{i}{2}\int dx \psi^*_\uparrow(x,t)\psi_\downarrow(x,t) \label{splus} \nonumber\\
&\equiv&\frac{ i}{2}|A(t)|e^{i\Delta \nu(t)t}\, .
\end{eqnarray}Here the Ramsey fringe contrast $\mathcal{C}(t) \equiv |A(t)|/|A(0)|$, is related to the overlap of the $\psi_\uparrow(x,t)$ and $\psi_\downarrow(x,t)$ wavefunctions. The decay of the contrast is a measure of demagnetization. The frequency  shift $\Delta \nu(t)$   measures the dynamical phase difference between the spins  and gives rise in this case  to a net motionally-induced precession of the total magnetization. Throughout we set $\hbar=1$.

\section{Linear Magnetic Field}\label{sec2}
First we consider the case of a magnetic field with linear spatial dependence $B_Z(x) = Bx/\Delta\mu$ where $B$ is a constant with units of { energy/length}. Adding a linear potential to a harmonic potential results in a new harmonic potential  with the same frequency but shifted in opposite directions for each of the two spin states. The potential is
\begin{eqnarray}
\hat{V}(x) = \frac{1}{2}m\omega^2 ( x +\hat{\sigma}^Z x_0)^2 - \frac{1}{2}m\omega^2x_0^2\,.
\end{eqnarray} Here, $a_H = \sqrt{1/m\omega}$ is the oscillator length and $x_0= \frac{B a_H^2}{\omega}$ is the displacement of the oscillator resulting from the magnetic field.

\begin{figure}[t]
\centering
\includegraphics[width=250pt]{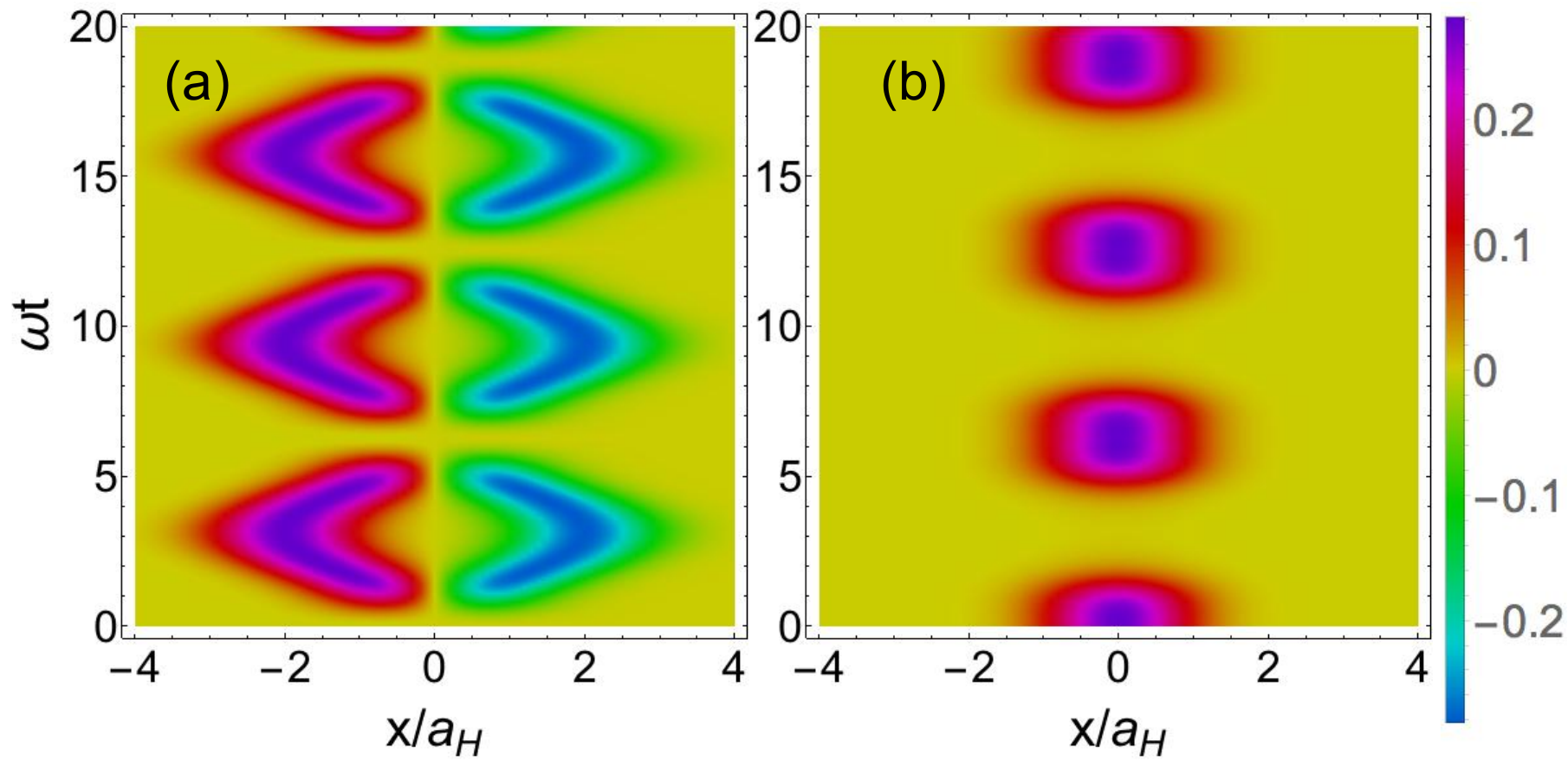}
\caption{Dynamics of spin observables in a linear magnetic field with $x_0=a_H$. (a) $ \langle \hat S^Z(x,t)\rangle=\frac{1}{2}\left(|\psi_\uparrow(x,t)|^2-|\psi_\downarrow(x,t)|^2\right)$ shows the spin up/down densities oscillating in their traps centered at $x=\mp x_0$. (b) Magnitude of $\langle \hat S^+(x,t)\rangle$, which measures the magnetization at point $x$. The magnetization decays when the spin up/down densities are separated. \label{linearheatplot}}
\end{figure}

It is well known that the displaced ground state of a harmonic oscillator evolves with a probability distribution of constant shape but oscillating centroid, $|\psi_0(x,t)|^2 \propto e^{-(x-x_0(1-\cos(\omega t))^2}$ \cite{sakurai}. A similar result can be derived for any eigenstate $n$. Namely, given a solution $\phi(x,t)$ to the time-dependent Schr\"odinger equation,
\begin{equation} \label{displacementformula} 
\psi(x,t) := \hat D\left(z_0 e^{-i\omega t}\right) \phi(x,t),
\end{equation}
is also a solution to the time-dependent Schr\"{o}dinger equation. The displacement operator is given by  $\hat D(w) \equiv \exp\left(w\hat a^\dagger - w^*\hat a \right)$ and $z_0$ is any complex number, corresponding to an initial displacement in the position-momentum  phase space \cite{sakurai}\footnote{Note that the displacement operator $\hat{D}(w)$ depends on the boson operators $\hat{a}$ which diagonalize the Hamiltonian at times $t>0$}.  This result allows us to calculate dynamics analytically and write the time evolution of an initial  eigenstate as (See Appendix A):
\begin{align}
\nonumber&\psi_n\left(x,t\right)=e^{-i \left(n+\frac{1}{2}\right)\omega t+i\frac{x_0^2}{a_H^2}\left(\frac{1}{2}\cos\left(\omega t\right)\sin\left(\omega t\right)-\sin\left(\omega t\right)\right)}\\
\label{eq:PRpsi}&\times e^{-i\sigma x_0 \sin\left(\omega t\right)x/a_H^2}\phi_n\left(\frac{x+\sigma x_0\left(1-\cos\left(\omega t\right)\right)}{a_H}\right)\,, 
\end{align}
where $\sigma$ is the eigenvalue of $\hat{\sigma}^Z$.  In Fig.~\ref{linearheatplot}(a), we plot the time evolution of the spin density in the $Z$-direction, $ \langle \hat S^Z(x,t)\rangle=\frac{1}{2}\left(|\psi_\uparrow(x,t)|^2-|\psi_\downarrow(x,t)|^2\right)$, for a single particle initially in the ground state of the harmonic oscillator, calculated using Eq. \ref{eq:PRpsi}. The $|\uparrow\rangle$ and $|\downarrow\rangle$ densities maintain their same shape but oscillate about their respective equilibrium positions at the trap frequency.

\subsection{Spin dynamics}

We now use the above results to determine the spin dynamics for a single particle.  Assuming that the particle begins in the $n^{\mathrm{th}}$ harmonic oscillator mode at time $t=0$, we find
\begin{eqnarray}
\label{SingleParticleSy}
&&\langle \hat S^+ \rangle = \frac{i}{2}  \exp\left(-2\frac{x_0^2}{a_H^2}\left(1 - \cos(\omega t)\right)\right) \nonumber \\
&&\times L_n\left(4\frac{x_0^2}{a_H^2}(1-\cos(\omega t))\right),
\end{eqnarray}
where $L_n$ is the $n^{\mathrm{th}}$ Laguerre polynomial.

We see that the contrast is suppressed exponentially with increasing $x_0$. This is caused by  the exponential suppression of the overlap integral between the spatial wavefunctions of the $\uparrow$ and $\downarrow$ spin states. As they move apart under the influence of their different potentials, the overlap decreases as a consequence of   the wavepackets' Gaussian spatial localization, see Fig.~\ref{linearheatplot}(b).
The contrast is periodic in time with period $2\pi / \omega$, due to the periodic motion of the two spin wavefunctions in the trap. We also see that there is no frequency shift for the case of a linear magnetic field and the direction of the magnetization remains along $Y$.  The two spin states, while displaced in position, experience harmonic potentials of the same frequency, so they have the same dynamical phase.

We now consider the dynamics of an ensemble of non-interacting fermions with this same protocol. An ensemble at zero temperature, initially spin-polarized, forms a Fermi-degenerate gas with all oscillator modes filled from $n=0$ to $n=N-1$, with $N$ the number of particles.  Here, we find
\begin{eqnarray}
&&\langle\hat S^+ \rangle^{\mathrm{f.d.}}=\displaystyle\sum\limits_{n=0}^{N-1} \langle \hat S^+ \rangle_n \nonumber\\
&&= \frac{i}{2} e^{-2\frac{x_0^2}{a_H^2}\left(1 - \cos(\omega t)\right)}\displaystyle\sum\limits_{n=0}^{N-1}L_n\left(4\frac{x_0^2}{a_H^2}(1-\cos(\omega t))\right) \nonumber\\
&&= \frac{i}{2} e^{-2\frac{x_0^2}{a_H^2}\left(1 - \cos(\omega t)\right)}L^1_{N-1}\left(4\frac{x_0^2}{a_H^2}(1-\cos(\omega t))\right), 
\end{eqnarray}
where $L^a_b(x)$ is the associated Laguerre polynomial and we have used the addition formula for Laguerre polynomials
$\sum_{m=0}^{n}L_m^{\left(a\right)}\left(x\right)L_{n-m}^{\left(b\right)}\left(y\right)=L_{n}^{\left(a+b+1\right)}\left(x+y\right)$. The contrast decay per particle at short times is given by 
\begin{eqnarray}
\mathcal{C}\left(t\right)/N=1-E_F\frac{x_0^2}{a_H^2}\omega t^2+\mathcal{O}\left(t^4\right),
\end{eqnarray}
 with $E_F=k_B T_F=N\omega$ the Fermi energy of the gas, $k_B$ is the Boltzmann constant, and $T_F$ the Fermi temperature.  Hence, the decay of the contrast per particle increases extensively as the number of particles increases.

For an ensemble at non-zero temperature,  the associated Fermi-Dirac sums cannot be evaluated exactly, but can be approximated by energy integrals with a constant density of states to yield a formula for the short-time contrast decay per particle:
\begin{eqnarray} \label{lincontrast}
&&\mathcal{C}\left(t\right)/N=1-2\bar{E}\frac{x_0^2}{a_H^2}\omega t^2+\mathcal{O}\left(t^4\right) \nonumber\\
&&\bar{E} \approx \frac{1}{6N\omega \beta^2}\left[\pi^2+3\ln(e^{\beta N\omega}-1)+6{\rm Li}_2\left(\frac{1}{1-e^{\beta N\omega}}\right)\right],\nonumber\\
\label{shorttimelinear}
\end{eqnarray}
where  $\beta = 1/k_BT$,  $T$ the temperature and ${\rm Li}_s\left(x\right)$ is the polylogarithm of order $s$.

In the limit of high temperature, $T/T_F\gg 1$, we can approximate the distribution function of the fermions with a Maxwell-Boltzmann distribution. For this ensemble the spin dynamics is given by
\begin{eqnarray}
&&\langle\hat S^+ \rangle^{\mathrm{M.B.}}=\displaystyle\sum\limits_{n=0}^\infty e^{-\beta\omega(n+1/2)}\langle \hat S^+ \rangle_n/\mathcal{Z} \nonumber\\
&&=\frac{i}{2} \exp\left[-2\frac{x_0^2\left(1-\cos(\omega t)\right)}{a_H^2\tanh(\beta\omega/2)}\right],
\end{eqnarray} where $\mathcal{Z}=\sum_ne^{-\beta\omega(n+1/2)}$ is the partition function.  At short times, we find the contrast per particle decays as $\mathcal{C}\left(t\right)/N=1-\frac{x_0^2}{a_H^2}\frac{1}{\tanh(\beta\omega/2)}\omega^2 t^2+\mathcal{O}\left(t^4\right)$, which agrees with Eq.~\eqref{lincontrast} when we identify the mean energy $\frac{1}{2}\tanh^{-1}(\beta\omega/2)=\bar{E}/\omega $.  At $T/T_F\gg 1$, where this analysis is valid, $\bar{E}\approx k_B T $, and so the contrast decay rate  increases linearly with temperature.

\subsection{Dynamics for a spin-echo sequence}
The spin-echo consists of an additional   collective $\pi$ rotation about $X$ added at time $t/2$.  This pulse swaps the states of the spin up and spin down components, with the goal of removing spin-dependent single-particle inhomogeneities.   It is natural to wonder whether a spin-echo pulse will remove the effects of a magnetic field gradient, which is effectively a spatially-inhomogeneous detuning, when the particles themselves move in the trap. In the presence of a spin-echo pulse the evolution is given by:
\begin{eqnarray}
&&\langle S^+\rangle_{\mathrm{SE}} = \frac{i}{2} \exp\left(-16 \frac{x_0^2}{a_H^2} \sin^4\left(\frac{\omega t}{4}\right)\right) \nonumber\\
&&\times L_n\left(32 \frac{x_0^2}{a_H^2} \sin^4\left(\frac{\omega t}{4}\right)\right)\, .
\end{eqnarray}
The spin dynamics with a spin-echo pulse is periodic with period $4\pi/\omega$, twice the period of a Ramsey sequence without a spin-echo pulse.  The short-time expansion of the contrast decay is
\begin{align}
\mathcal{C}_{\mathrm{SE}}\left(t\right) &= 1-\frac{1}{8}\left(n+\frac{1}{2}\right) \frac{x_0^2}{a_H^2}(\omega t)^4+\mathcal{O}\left(t^6\right)\,.
\end{align}
Note that the spin-echo removes the dominant $\mathcal{O}\left(t^2\right)$ contribution to the contrast decay at times much shorter than   the motional period.  This is consistent with the expectation that spin-echo removes the effect of inhomogeneities when  motional effects can be neglected ~\cite{Carr_Purcell_54}. Beyond the short time limit the spin-echo pulse is not beneficial.

\begin{figure}
\centering
\includegraphics[width=200pt]{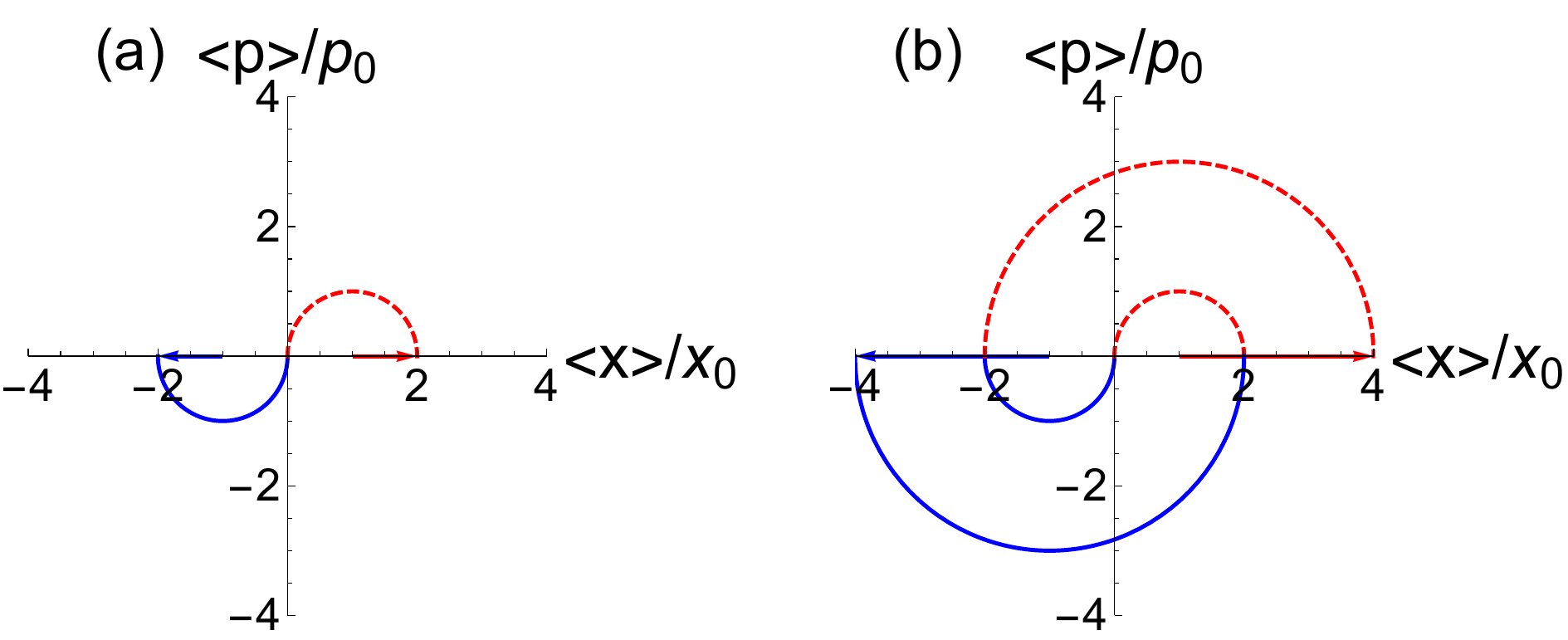}
\caption{Phase space plots of $\langle x \rangle$ and $\langle p \rangle$ for the $|\uparrow\rangle$ state (blue, solid) and $|\downarrow\rangle$ state (red, dashed) for a spin-echo sequence in a linear magnetic field. Here $ p_0=1/a_0$, and the total time is $2\pi/\omega$. The two spin states evolve along circular trajectories centered at $\mp x_0$. (a) At half the total dark time, $\pi/\omega$, the spin-echo $\pi$ pulse swaps the spin states. The spin wavefunctions are thus displaced by $2x_0$ from the centers of their respective harmonic traps. For the second half of the dark time they evolve in circular trajectories twice as large, centered at $\mp 2x_0$, resulting in a final configuration shown in (b). The spin-echo pulse  leads to greater dephasing of the spin observables due to larger final phase-space displacement.\label{spinechophasespace}}
\end{figure}

The behavior of the spin-echo signal at times comparable to the trap period can be visualized in phase space.  To illustrate the dynamics, we rescale the phase space coordinates to $x/x_0$ and $p/p_0$, where $p_0=1/a_0$, see Fig. (\ref{spinechophasespace}).  In this coordinate system, the expectation values $\langle x \rangle$ and $\langle p \rangle$ for the $|\uparrow\rangle$ state and $|\downarrow\rangle$ states initially move along circular trajectories centered at $\mp x_0$, reflecting their initial displacement from the center of their respective harmonic traps due to the magnetic field.  The spin-echo $\pi$ pulse switches the $|\uparrow\rangle$ and $|\downarrow\rangle$ states and while doing so it generally   enhances  the net  displacement of the spin states from their motional centers. After the pulse the  dynamics corresponds to  circular trajectories with a new phase-space  radius. For  example, as shown in Fig.~\ref{spinechophasespace},  if the echo is applied at  $\pi/\omega$  (dark time $t=2\pi/\omega$), the displacements following the spin-echo pulse are twice the displacements before the pulse.

\begin{figure}
\centering
\includegraphics[width=200pt]{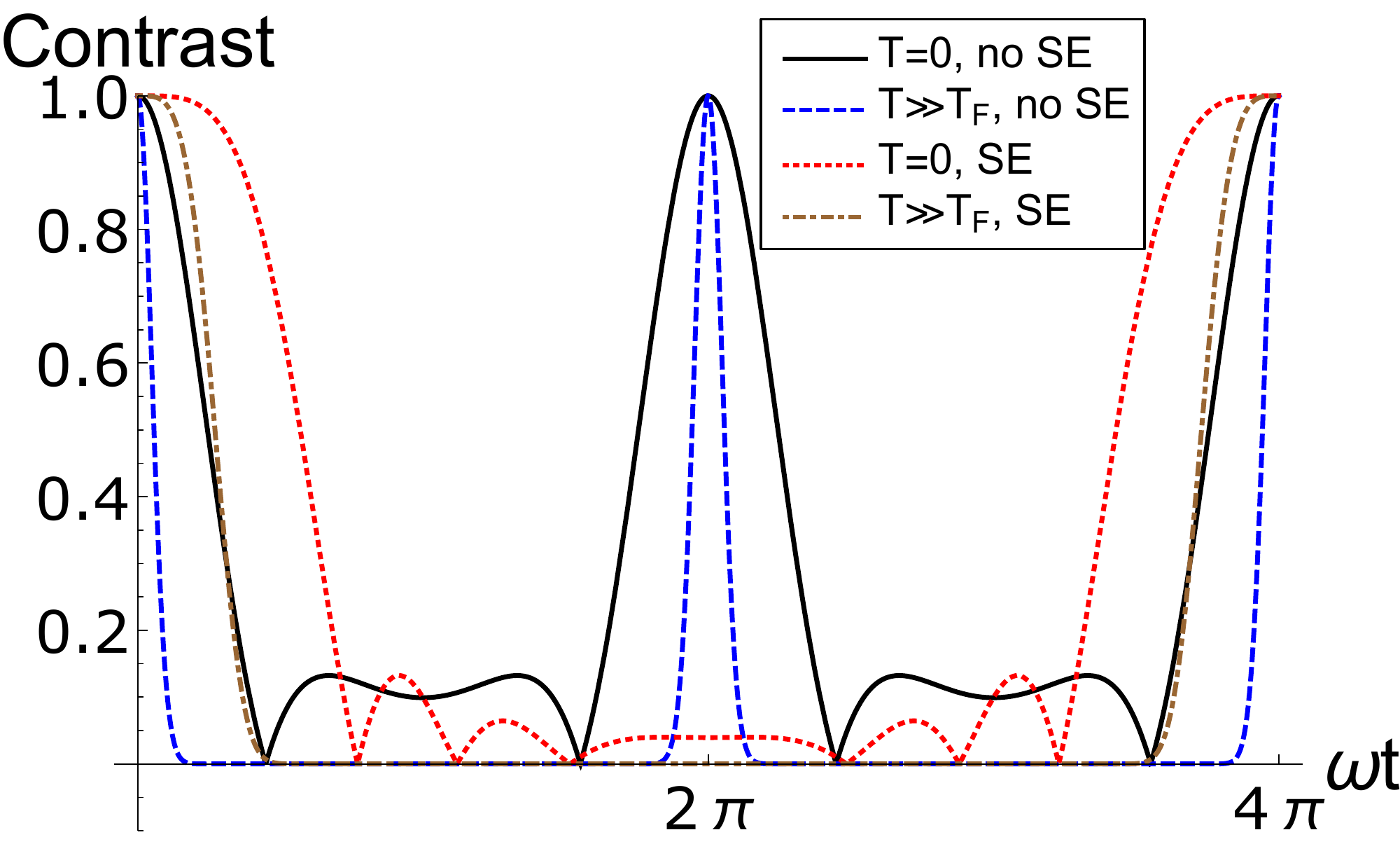}
\caption{Contrast decay for the case of a linear magnetic field at times comparable to the trap period, for zero temperature and high temperature ($T \gg T_F$), with and without spin-echo (SE).  The quantum degenerate case uses $N=19$ and all cases use $x_0=0.24a_H$, consistent with Ref. \cite{Bardon14}. $T=10T_F$ in the high temperature case. Higher temperature leads to faster contrast decay since the particles have a larger average energy. The spin-echo removes the second-order contribution to the contrast decay, resulting in slower decay at short times. However, at long times the contrast decay is larger when a spin-echo pulse is applied. The period of the dynamics is also $4\pi/\omega$ with a spin-echo pulse instead of $2\pi/\omega$. \label{therm_contrast}}
\end{figure}

 For dark times less than half the motional period, the spin-echo improves the contrast, consistent with the short-time analysis. In the phase space picture this means that  the displacement between $|\uparrow \rangle$ and $|\downarrow \rangle$ with spin-echo is smaller than without spin-echo. At a dark time of exactly half the motional period, the contrast decay with and without spin-echo is identical -- this is illustrated in Fig. \ref{therm_contrast}. In this case the phase space displacement without spin-echo is purely along $x$ and the displacement with spin-echo is purely along $p$. For longer dark times, spin-echo increases dephasing, since it produces larger phase space displacements. The dephasing is maximal at dark times which are odd-integer multiples of the motional period. When the spin-echo pulse is applied at an integer multiple of the motional period, i.e. dark times of integer multiples of twice the motional period $t=4\pi k/\omega$ ($k$ an integer), both spin states have returned to the phase-space origin and the motional dynamics is unaffected by the spin-echo pulse.  This explains the periodicity of the spin-echo signal with twice the motional period.  We stress that the ability to decouple the spin-echo effect from motion is due to the two spin states sharing a common motional frequency.  Although spin-echo is typically used to eliminate dephasing due to single particle inhomogeneities, in the presence of spin-motional coupling the particle motion  during the dark time can worsen the resulting dephasing when  a spin-echo pulse is applied.

We can calculate the dynamics of thermal ensembles with a spin-echo pulse.  At zero temperature, the exact result is
\begin{eqnarray}
&&\langle\hat S^+ \rangle^{\mathrm{f.d.}}_{\mathrm{SE}} = \frac{i}{2}\exp\left(-16 \frac{x_0^2}{a_H^2} \sin^4\left(\frac{\omega t}{4}\right)\right) \nonumber\\
&&\times L_{N-1}^1\left(32 \frac{x_0^2}{a_H^2} \sin^4\left(\frac{\omega t}{4}\right)\right)\, .
\end{eqnarray}
For an ensemble at arbitrary temperature the short time contrast decay is given by
\begin{eqnarray}
&&\mathcal{C}_{SE}\left(t\right)/N=1-32\bar{E}\frac{x_0^2}{a_H^2}\omega^3 t^4+\mathcal{O}\left(t^6\right) \nonumber\\
&&\bar{E} \approx \frac{1}{6N\omega \beta^2}\left[\pi^2+3\ln(e^{\beta N\omega}-1)+6{\rm Li}_2\left(\frac{1}{1-e^{\beta N\omega}}\right)\right]\, ,\nonumber \\ \label{shorttimeSElinear}
\end{eqnarray}
and at high temperature we can approximate the ensemble by a Maxwell-Boltzmann distribution to obtain
\begin{eqnarray}
&&\langle\hat S^+ \rangle^{\mathrm{M.B.}}_{\mathrm{SE}}=
\frac{i}{2}\exp\left[-16\frac{x_0^2\sin^4\left(\frac{\omega t}{4}\right)}{a_H^2\tanh(\beta\omega/2)}\right]\, .
\end{eqnarray}
The contrast at zero temperature and high temperature ($T\gg T_F$) are shown with and without spin-echo as a function of dark time in Fig.~\ref{therm_contrast}. 

\begin{figure}
\centering
\includegraphics[width=200pt]{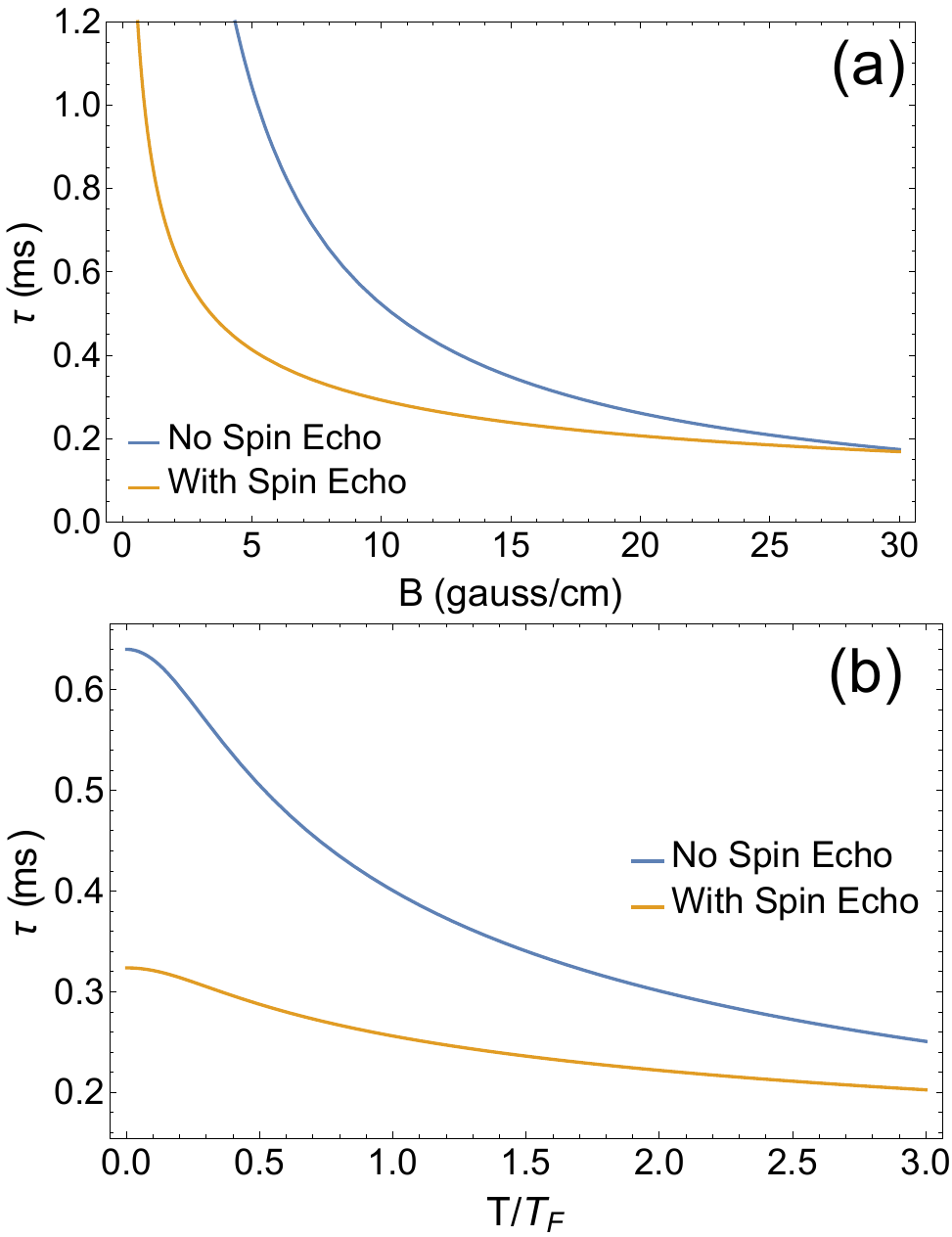} 
\caption{Linear magnetic field: (a) demagnetization time vs. magnetic field gradient. (b) Demagnetization time vs. temperature where the Fermi temperature $T_F = N\omega$. Using parameters from Ref. \cite{Bardon14}, these timescales are faster than those obtained in the experiment, suggesting that strong interactions suppress the role of single particle dynamics.  \label{sp_linear_figure}}
\end{figure}

\subsection{General Considerations}
\label{sec:SpinDiff}

In general terms the contrast dynamics in the non-interacting regime is characterize by oscillatory behavior  at the  trap period arising  from spin-motion coupling. At short times the contrast decays as $C(t) \approx 1-\left(\frac{t}{\tau}\right)^2$ and in the presence of an echo-pulse as $C^{\mathrm{SE}}(t) \approx 1-\left(\frac{t}{\tau^{\mathrm{SE}}}\right)^4$, where we identify $\tau$ and $\tau^{\mathrm{SE}}$ as the demagnetization timescales without and with spin echo. Using Eqs. \ref{shorttimelinear} and \ref{shorttimeSElinear}, the demagnetization timescales are given by:
\begin{eqnarray}
&&\tau_{\mathrm{lin}}=\frac{a}{x_0}\left(2\omega\bar{E}\right)^{-1/2} \nonumber\\
&&\tau_{\mathrm{lin}}^{\mathrm{SE}}=\left(32\omega^3\frac{x_0^2}{a_H^2}\bar{E}\right)^{-1/4} \nonumber\\
&&\bar{E}\approx \frac{1}{6N\omega \beta^2}\left[\pi^2+3\ln(e^{\beta N\omega}-1)+6{\rm Li}_2\left(\frac{1}{1-e^{\beta N\omega}}\right)\right]\, .\nonumber \\
\end{eqnarray}The short time demagnetization rate  can be significantly  faster than demagnetization rates measured in recent experiments performed at unitarity \cite{Koschorreck2013,Bardon14,Trotzky15}, where the magnetization decay is manly diffusive with a short time scaling given by  $
 C^{\mathrm{U}}(t) \approx 1-\left(\frac{t}{\tau^{\mathrm{U}}}\right)^3$.
 The different time dependence  translates into a different dependence on the magnetic field gradient: while in the  non-interacting regime the  demagnetization rate  scales as $1/\tau \sim B$ and $ 1/\tau^{SE} \sim \sqrt{B}$,  in the interaction dominated regime it scales as  $1/\tau \sim B^{2/3}$ \cite{Koschorreck2013,Bardon14,Trotzky15}.

The demagnetization rate in the non-interacting system increases with temperature at  fixed particle number with and without echo. 
The increasing decay rate in the non-interacting limit  simply reflects  the increase in the mean energy  per particle as the gas gets hotter. This behavior persists in the presence of interactions as observed in current experiments  given that  both the typical velocity and the mean free path (parameters that  determine the   spin diffusivity)  increase with temperature. Fig. (\ref{sp_linear_figure}) (a-b) shows the demagnetization timescales. We use an effective one-dimensional number of particles, which scales like $N^{1/3}$, where $N$ is the total number of particles in a three dimensional trap. The fact that the demagnetization rate due to the motion of the atoms in the trap can be comparable or to faster than the  spin diffusion decay rate makes it  clear that one relevant role of  strong interactions is to suppress the effects of single-particle motion.

\section{Quadratic Magnetic Field}\label{sec3}
We now consider the case when the applied inhomogeneous magnetic field varies quadratically with position: $B_z(x)=mBx^2/\Delta\mu$. Here $B$ is a constant with units of frequency$^2$. The total potential experienced by the atoms is
\begin{eqnarray}
\hat{V}(x)= \frac{1}{2}m\left(\omega^2 + \hat{\sigma}^Z B\right)x^2\, .\end{eqnarray}
Thus, each spin state sees a new trap frequency $\omega_\sigma = \sqrt{\omega^2 +\sigma B},$ with $\sigma$ the eigenvalue of $\hat{\sigma}^Z$.

We can find the time dynamics of a harmonic oscillator eigenstate subjected to a sudden quench in the oscillator frequency by making the following ansatz, analogous to Eq.~\eqref{eq:PRpsi} for the linear case:
\begin{align}
\label{eq:quadexact}\psi_n\left(x,t\right)&=\frac{e^{-i\left(n+\frac{1}{2}\right)\phi_{\sigma}\left(t\right)}}{\sqrt{a_H \alpha_{\sigma}\left(t\right)}}\psi_n\left(\frac{x}{a_H\alpha_{\sigma}\left(t\right)}\right)e^{i\beta_{\sigma}\left(t\right) \frac{x^2}{2a_H^2}}\, ,
\end{align}
where $\alpha_{\sigma}\left(t\right)$, $\beta_{\sigma}\left(t\right)$, and $\phi_{\sigma}\left(t\right)$ real functions which are independent of $n$. It can be shown that this ansatz represents a solution to the time-dependent Schr\"{o}dinger equation if the functions $\alpha_{\sigma}\left(t\right)$, $\beta_{\sigma}\left(t\right)$, and $\phi_{\sigma}\left(t\right)$ satisfy
\begin{eqnarray}
&&\label{ermakov}\alpha_{\sigma}'' \left(t\right)+\left(\omega^2+\sigma B\right)\alpha_{\sigma}\left(t\right)=\frac{\omega^2}{\alpha_{\sigma}^3\left(t\right)}, \nonumber \\
&&\beta_{\sigma}\left(t\right)=\frac{\alpha_{\sigma}'\left(t\right)}{\alpha_{\sigma}\left(t\right)}, \nonumber \\
&&\phi_{\sigma} '\left(t\right)=\frac{1}{\alpha_{\sigma}^2\left(t\right)}.
\end{eqnarray}
Equation~\eqref{ermakov} is a special case of the Ermakov equation, a well-studied nonlinear ordinary differential equation \cite{Pinney50}.
Defining $\tilde B = B/\omega^2$ as a dimensionless parameter, the solutions for our particular case are
\begin{align}
\label{eq:alphadef} \alpha_{\sigma}\left(t\right)&=\sqrt{\frac{2+\tilde B\sigma\left(1+\cos\left(2\omega t\sqrt{1+\tilde B\sigma}\right)\right)}{2\left(1+\tilde B\sigma\right)}}, \nonumber \\
 \beta_{\sigma}\left(t\right)&=-\frac{\tilde B \sqrt{1+\tilde B\sigma}\sin\left(2\omega t\sqrt{1+\tilde B\sigma}\right)}{2+\tilde B\sigma\left(1+\cos\left(2\omega t\sqrt{1+\tilde B\sigma}\right)\right)},\, \nonumber \\
\phi_{\sigma}\left(t\right)&=\lfloor\frac{\omega t \sqrt{1+\tilde B}\sigma}{\pi}\rfloor \pi +\nonumber \\
+\frac{\pi}{2}-&\arctan\left[\sqrt{1+\tilde B\sigma} \cot\left(\omega t\sqrt{1+\tilde B\sigma}\right)\right]\, ,
\end{align}
where $\lfloor . \rfloor$ is the floor function.
\begin{figure}
\centering
\includegraphics[width=250pt]{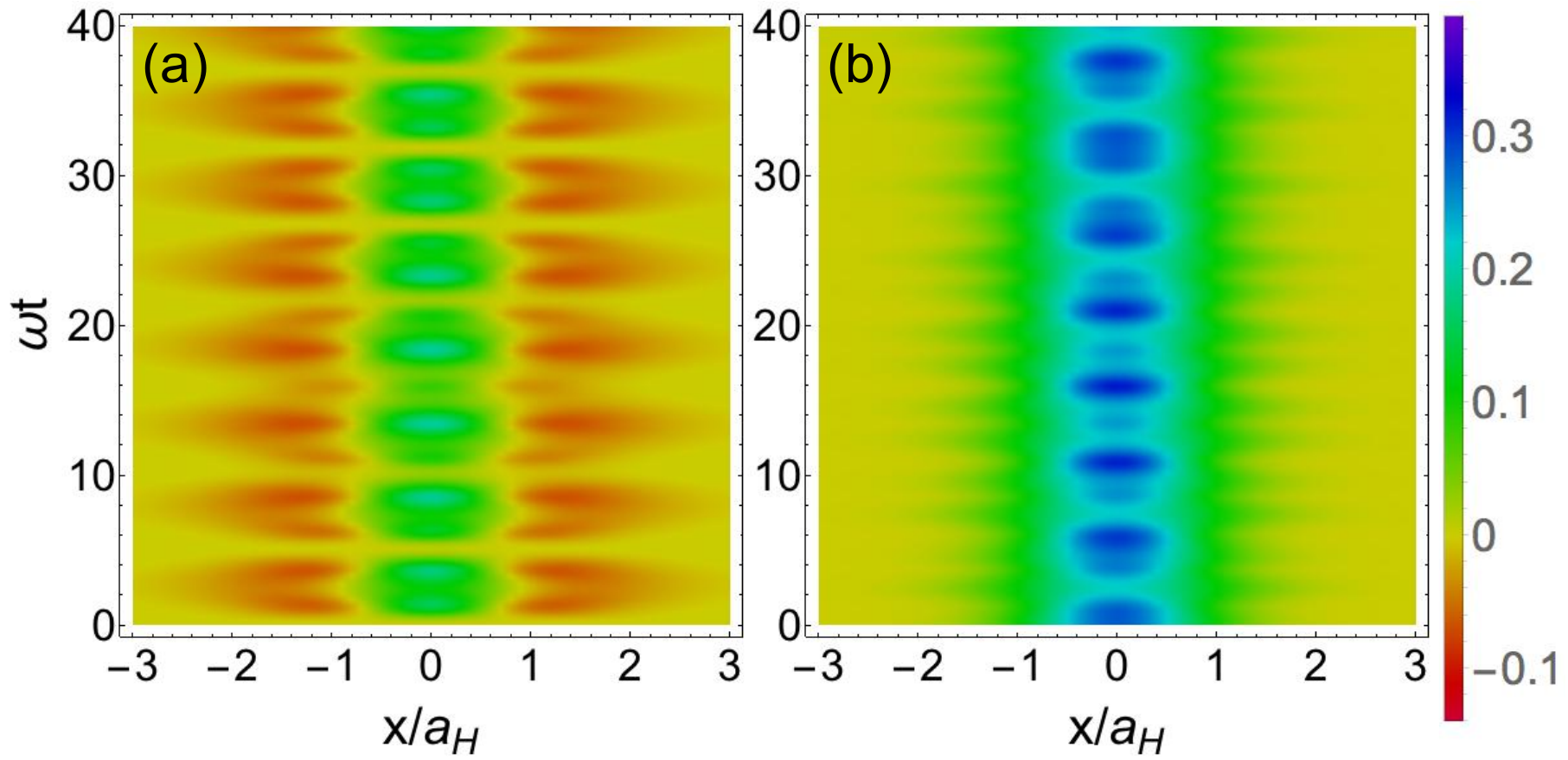}
\caption{ Dynamics of spin observables in a quadratic magnetic field with $B=0.65\omega^2$. (a) $\langle \hat  S^Z(x,t) \rangle$ shows the spin up/down densities breathing in their traps centered at $x=0$. The rates of their breathing, $\omega^{\pm}=\sqrt{\omega^2\pm B}$, are incommensurate. The standard deviations  of the densities are $a^\pm = 1/\sqrt{\omega^\pm}$, hence spin up (down) becomes more concentrated at the center (edge) of the trap. (b) Magnitude of $\langle \hat S^+(x,t)\rangle$ (contrast), which measures the magnetization at position $x$. The magnetization decays when the spin up/down densities are separated. Note that decays/revivals of the magnetization are aperiodic.} \label{quadheatplot}
\end{figure}
The solutions $\psi_n\left(x,t\right)$ are wavefunctions whose probability densities maintain their original shape, up to periodic rescaling by $\alpha(t)$.  The frequency associated with the ``breathing'' of the probability density is $2\omega\sqrt{1+ \tilde B\sigma}$, and hence the periods of oscillation are generally incommensurate for the spin up and spin down particles, see Fig.~(\ref{quadheatplot}).  These modes are the breathing analogs of the well-known coherent states of a harmonic oscillator.

 Figure~(\ref{quadheatplot}) also shows

 \begin{equation}
 \langle \hat  S^Z(x,t)\rangle =\frac{1}{2}\left(|\psi_\uparrow(x,t)|^2-|\psi_\downarrow(x,t)|^2\right),\end{equation}   that reflects the breathing of the  spin up/down densities  around  the trap centers at $x=0$.  The different standard deviations  of the densities,  $a^\pm = 1/\sqrt{\omega^\pm}$, also lead to a spin segregation, or spatial separation of the spin up/down densities: while the former  becomes more localized towards the center, the latter becomes more concentrated at the edge of the trap.

\subsection{Spin Dynamics}

To calculate $\langle \hat S^+\rangle = \frac{i}{2}\int dx \psi^*_\uparrow(x,t)\psi_\downarrow(x,t)$ we use Eq.~\eqref{eq:quadexact} and write
\begin{eqnarray} \label{Hnint}
\langle \hat S^+\rangle = \frac{i}{2}\frac{\exp\left[i(\phi_{\uparrow}(t)-\phi_{\downarrow}(t))\right]}{\sqrt{ \alpha_{\uparrow}(t)\alpha_{\downarrow}(t)}}I_n ,
\end{eqnarray}
where
\begin{align}
I_n&\equiv \frac{1}{2^n n!\sqrt{\pi} }\int dx e^{-ax^2}H_n\left(bx\right)H_n\left(cx\right) \nonumber \\
a&=\frac{\alpha_{\uparrow}(t)^2+\alpha_{\downarrow}(t)^2}{2\alpha_{\uparrow}(t)^2\alpha_{\downarrow}(t)^2}+\frac{i}{2}\left(\beta_{\uparrow}(t)-\beta_{\downarrow}(t)\right) \nonumber \\
b&=\frac{1}{\alpha_{\uparrow}(t)}, \qquad c=\frac{1}{\alpha_{\downarrow}(t)}\, ,
\end{align}
and $H_n(x)$ is a Hermite polynomial.  Due to the difference in breathing frequencies, the spin up and spin down particles accumulate a different dynamical phase, unlike the case of a linear magnetic field.  Because of this dynamical phase difference, there is a net frequency shift.

Using the generating function for the Hermite polynomials~\cite{Gottfried_Yan}, we can find a generating function for the integrals $I_n$ as
\begin{align}
g_I\left(z\right)\equiv \frac{1}{\sqrt{a-2bcz+\left(b^2+c^2-a\right)z^2}}=\sum_{n=0}^{\infty} I_n z^n\, ,
\end{align}
which leads to the following closed form for $\langle \hat S^+\rangle$:
\begin{eqnarray}
&&\langle \hat S^+\rangle = \frac{i}{2} \frac{e^{i( \phi_{\uparrow}(t)-\phi_{\downarrow}(t))}}{\sqrt{a\alpha_{\uparrow}(t)\alpha_{\downarrow}(t)}}\left(g_1 g_2\right)^n \times \nonumber \\
&&\times _2F_1\left[\frac{1-n}{2},\frac{-n}{2},1,\frac{(g_1^2-1)(g_2^2-1)}{g_1^2g_2^2}\right]\, .
\end{eqnarray}
Here, $g_1=b/\sqrt{a}$, $g_2=c/\sqrt{a}$, and $_2F_1\left[.\right]$ is the hypergeometric function. To illuminate this result we can expand $\langle \hat S^+\rangle$ to second order in $\omega t$:
\begin{eqnarray}
\label{eq:SpshortTimes}&&\langle \hat S^+\rangle \approx \frac{i}{2} \Big[1+\frac{iB}{\omega^2}\left(\omega t\right)(n+\frac{1}{2}) +\nonumber \\
&&-\frac{3}{4}\left(\frac{B}{\omega^2}\right)^2\left(\omega t\right)^2\left(n^2+n+\frac{1}{2}\right) + O\left(\left( \omega t\right)^3\right)\Big].
\end{eqnarray}
We see that at short times the signal acquires a frequency shift $\Delta \nu = (B/\omega)(n+\frac{1}{2})$ and thus the collective spin exhibits a net precession in the $X$-$Y$ plane. The frequency shift is a consequence of the additional frequency scale in the problem, namely, $\omega_\uparrow - \omega_\downarrow = \sqrt{\omega^2+B}-\sqrt{\omega^2-B} \approx B/\omega$ for $B \ll \omega^2$. Additionally, the contrast decays quadratically in $Bt/\omega$, and depends on the mean of the squared energy.  From the generating function, we can also find the Maxwell-Boltzmann thermal average exactly as $g_I\left(e^{-\beta \omega}\right)/\mathcal{Z}$, which gives
\begin{widetext}
\begin{align}
\label{eq:quadMB} \langle \hat S^+\rangle^{\mathrm{M.B.}}&=\frac{i\exp\left[i(\phi_{\uparrow}(t)-\phi_{\downarrow}(t))\right] \sinh\left({\beta\omega }/{2}\right)}{\sqrt{(\alpha^2_{\downarrow}(t)+\alpha^2_{\uparrow}(t))\cosh(\beta\omega)+\alpha_{\downarrow}(t)\alpha_{\uparrow}(t)(i\alpha_{\downarrow}(t)\alpha_{\uparrow}(t)(\beta_{\uparrow}(t)-\beta_{\downarrow}(t))\sinh(\beta\omega)-2)}}\, .
\end{align}
\end{widetext}


\subsection{Spin-echo sequence for quadratic magnetic field}
To find the spin dynamics we solve Ermakov equation with the conditions of a spin-echo sequence.
The result, expanded at short times, is
\begin{eqnarray}
&&\langle \hat S^+\rangle \approx \frac{i}{2}\Big[1-\frac{i}{8}\left(\frac{B}{\omega^2}\right)\left(\omega t\right)^3(n+\frac{1}{2}) + \nonumber \\
&&-\frac{1}{16}\left(\frac{B}{\omega^2}\right)^2\left(\omega t\right)^4\left(n^2+n+1\right) + O\left(\left(\omega t\right)^5\right)\Big]. \label{eq:SpshortTimes}
\end{eqnarray}
As was the case for the linear magnetic field, the spin-echo removes the leading order contrast decay, and also removes the leading order frequency shift.  We again interpret this result as due to the fact that the particles do not move to first order in time, and so the spin-echo can remove the effectively-static dephasing at lowest order.  For longer times, there is an essential difference in the present case compared to the magnetic field with linear position dependence:  here, $\langle \hat S^+\rangle$ is not a periodic function of time, as the two frequencies $ \sqrt{\omega^2+B}$ and $\sqrt{\omega^2-B}$ are incommensurate for general $B$.  In the case of a linear magnetic field, the residual effects of the spin-echo pulse on motion can be removed by applying the spin-echo pulse after a single motional period, as both spin states oscillate at the trap frequency.  In the quadratic case, the two spin states have different motional periods, and so a spin-echo pulse cannot remove motional effects from both spin states at long times.

\begin{figure}
\centering
\includegraphics[width=200pt]{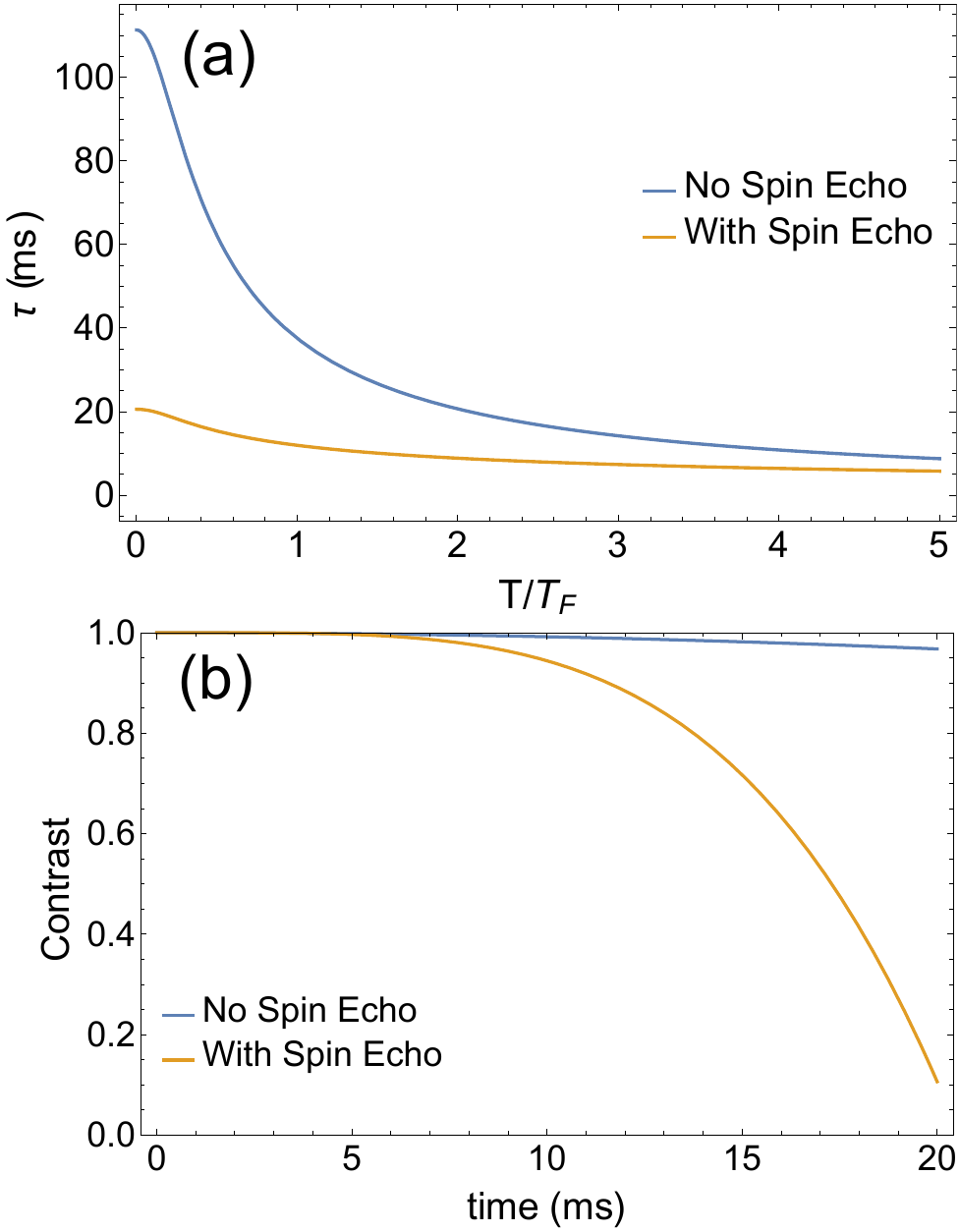}
\caption{ Quadratic magnetic field: (a) demagnetization time with and without spin-echo vs. temperature, where the Fermi temperature $T_F = N\omega$, and (b) contrast decay vs. time, using parameters from Ref.  \cite{Du2008}. The timescales are on the order of or faster than the timescale of spin segregation in the experiment $(\sim 100 {\rm ms})$. Additionally, application of a spin-echo pulse leads to faster demagnetization.} \label{sp_quad_figure}
\end{figure}

\subsection{General Considerations}

For the quadratic magnetic field  the contrast decay due to motional effects, expressed as a demagnetization rate, also exhibits a quadratic and quartic scaling with time with and without a spin-echo pulse, respectively. However, unlike the linear magnetic field case, there are generally no full revivals of the magnetization, due to the incommensurate breathing frequencies of the spin up and spin down densities.
Using the approximation of a continuum of energies and a constant density of states, the corresponding demagnetization timescales are given by:
\begin{eqnarray}
&&\tau_{\mathrm{quad}}=\frac{2 }{\sqrt{3}}\left(\frac{\omega^2}{B}\right)\left(\bar{E^2}-\omega^2/2\right)^{-1/2}\nonumber\\ \nonumber
&&\tau^{\mathrm{SE}}_{\mathrm{quad}} =2\sqrt{\frac{\omega^2}{B}}\left[\omega^2\left(\bar{E^2}-3\omega^2/4\right)\right]^{-1/4}\\
&&\bar{E}^2 \approx \frac{\pi^2\ln(e^{\beta N\omega}-1)+\ln(e^{\beta N\omega}-1)^3- 6{\rm Li}_3\left(\frac{1}{1-e^{\beta N\omega}}\right)}{3\beta^3\omega(N+1)}\, ,\nonumber\\
\end{eqnarray}
where the last line gives the mean squared energy.  In comparison to the case of a constant gradient, the rate is proportional to the mean squared energy rather than the mean energy, so the demagnetization rates increase more quickly with particle number and temperature than the linear case.  In Fig.~\ref{sp_quad_figure} we plot the demagnetization time and contrast decay for the quadratic magnetic field case with and without spin-echo, using experimental parameters from Ref. \cite{Du2008}, but with the proper rescaling of particle number to account for the different dimensionality,  as was done for the linear magnetic field case.   Under these conditions  a spin-echo pulse  again leads to faster demagnetization. We find that the demagnetization  timescales are on the order of or faster than the spin segregation timescales observed in an experiment in which interactions were non-negligible $(\sim 100 {\rm ms})$. There, the  segregation timescales were dictated  by  the mean interaction energy of the gas.  The faster short-time decay exhibited by the non-interacting system reveals once more the suppression of motional effects from interactions.
\section{Summary and Outlook}\label{sec4}
We have demonstrated that the motion of non-interacting fermions in a trap can lead to demagnetization on timescales  faster than  those caused by interactions in recent experiments. Additionally, a spin-echo sequence can have the counterintuitive effect of enhancing the rate of demagnetization at times comparable to or longer than the motional period. The analysis present here, which exactly characterizes all the relevant timescales and parameters that determine the non-interacting spin dynamics of a finite temperature fermionic  gas, can serve as a  platform for a better understanding of the  interplay between motional-induced and interaction-induced spin dynamics.

A logical direction for future work is to characterize the crossover from non-interacting to interacting dynamics.  A promising avenue along these lines is to treat the occupation of single-particle energy states as fixed and consider the effect of interactions on these occupied single-particle states. This will be explored in detail in an upcoming work.

During the finalization of this work we became aware of a complementary paper that treats the linear magnetic field case at zero temperature (Ref. \cite{xu2015}).\\

\section{Acknowledgements}
This work was supported by JILA-NSF-PFC-1125844, NSF-PIF-
1211914, ARO, AFOSR, and AFOSR-MURI. AK was supported by the Department of Defense through the NDSEG program.  MLW thanks the NRC postdoctoral fellowship program for support.

\newpage
\bibliography{spinsegcitations}

\appendix
\section{Derivation of generalized coherent-state formula}
Consider $\hat H = \frac{\hat p^2}{2m} + \frac{m\omega^2 \hat x^2}{2} = \omega\left(\hat a^\dagger \hat a + \frac{1}{2}\right)$ and suppose that $|\phi\rangle$ is a solution to the (time-dependent) Schr\"odinger equation. We now show that $|\psi\rangle :=\hat D\left(z := z_0e^{-i\omega t}\right)|\phi\rangle$ is also a solution, where $\hat D$ is the phase-space displacement operator $\hat D(w) = \exp\left(w\hat a^\dagger - w^*\hat a \right)$ and $z_0$ is any complex number, corresponding to the initial displacement in phase space.

Observing that $[\hat a,\hat D(z)] = z\hat D(z)$ and $[\hat a^\dagger,\hat D(z)] = z^*\hat D(z)$, we have
\begin{equation}\begin{split}
[\hat H, \hat D(z)] &=  \omega \left(z\hat a^\dagger \hat D(z) + z^*\hat D(z)\hat a\right) \\
&= \omega \left(z\hat a^\dagger + z^* \hat a - z^*z\right)\hat D(z).
\end{split}\end{equation}

Noting also that $\dot z = -i \omega z, \dot z^* = i \omega z^*$, we have from the Baker-Campbell-Hausdorff formula,
\begin{equation}\begin{split}
\frac{\partial}{\partial t} \hat D(z) &= \frac{\partial}{\partial t} \left(e^{z\hat a^\dagger} e^{-z^*\hat a} e^{-z_0^*z_0/2}\right) \\
&= -i \omega z \hat a^\dagger \hat D(z) + e^{z\hat a^\dagger} (-i \omega z^* \hat a) e^{-z^*\hat a} e^{-z_0^*z_0/2} \\
&= \left( - i \omega z \hat a^\dagger - i \omega z^* \hat a + i \omega z^* z\right) \hat D(z),
\end{split}\end{equation}
since $[e^{z\hat a^\dagger}, \hat a] = -ze^{z\hat a^\dagger}$. Thus 
\begin{equation}\begin{split}
i \frac{\partial |\psi\rangle}{\partial t} &= \omega \left(z\hat a^\dagger + z^* \hat a - z^*z\right)\hat D(z)|\phi\rangle + \hat D(z) \left(i \frac{\partial |\phi\rangle}{\partial t}\right) \\
&= [\hat H, \hat D(z)] |\phi\rangle + \hat D(z) \hat H |\phi\rangle \\
&= \hat H \hat D(z) |\phi\rangle = \hat H |\psi\rangle.
\end{split}\end{equation}

\end{document}